# The Existence of the Tau One-Way Functions Class as a Proof that P ≠ NP

Javier A. Arroyo-Figueroa

Entevia, LLC[1][2]

email: jarroyo@entevia.com

## Abstract

We prove that P ≠ NP by proving the existence of a class of functions we call T (Greek Tau), each of whose members satisfies the conditions of one-way functions. Each member of T is a function computable in polynomial time, with negligible probability of finding its inverse by any polynomial probabilistic algorithm. We also prove that no polynomial-time algorithm exists to compute the inverse of members of T, and that the problem of computing the inverse of T cannot be reduced to FSAT in polynomial time.

## 1. Introduction

This paper presents a proof that P ≠ NP by proving of the existence of a class of functions we call T (Greek Tau), each of whose members satisfies the conditions of one-way functions. Each member of T is a function computable in polynomial time, with negligible probability of finding its inverse by any polynomial-time probabilistic algorithm. This is accomplished by constructing each $\tau \in T$ with a collection of independent universal hash functions that produce a starting coordinate and path within a sequence of unique random bit matrices.

The existence of one-way functions has been an open question in computer science. This is due to the fact that, given a candidate function, it is hard to prove that any polynomial-time random algorithm that attempts to find any member of the inverse of the candidate, has a negligible probability of success, where such

---

probability is a function of the size of the input. To prove this, it is important to prove that no algorithm that is able to find the inverse of the candidate function runs in polynomial time; otherwise, such algorithm can be simulated by a probabilistic polynomial-time (PPT) algorithm with full certainty. Also, the proof of the non-existence of such algorithm in FP makes our proof not to rely on the assumption of the resolution of the P vs. NP problem (which would be a recursive, non-constructive proof).

Since every solution of the inverse of a one-way function can be verified in polynomial time (the FNP class), the existence of a one-way function means that there are problems in FNP that do not have a polynomial-time algorithm (the FP class) for finding the inverse. Thus, the existence of one-way functions implies that FP ≠ FNP, and thus P ≠ NP.

The rest of this paper is organized as follows. In the next section, we present the foundational definitions used in our proofs. In Section 3, the Tau functions class is presented. A survey of related works is presented in Section 4. A definition of the T-INV and H-INV problems are presented in sections 5 and 6, respectively. In Section 7, we prove that some instances of T-INV cannot be reduced to FSAT in polynomial time. In Section 8, we present lower bounds for the worst-time complexity of T-INV, followed in Section 9 by a proof of the one-wayness of all members of T. Our conclusions are presented in Section 10.

## 2. Definitions

A universal hash family H is defined as a set of hash functions

$$H = \{ h : U \to M \mid U \in \{0,1\}^n, M \in \{0,1,\cdots,m\}, m < 2^n \}$$

that satisfies the following property:

$$\forall x,y \in U, x \neq y \quad \Pr_{h \in U} [ h(x) = h(y) ] \leq 1/m$$

which means that any two members of U collide with probability at most $1/m$. Universal hash functions provide the guarantee that, for any $S \subseteq U$, and for any $x,y \in S$, the expected number of values of y that satisfy $h(x)=h(y)$ is $n/m$ [CW77].

A function f : {0,1}* → {0,1}* is *one-way* if and only if it can be computed in polynomial time, but any probabilistic polynomial time (PPT) algorithm F⁻ will only succeed in finding the inverse of f with *negligible* probability. More formally, f is one-way iff:

$$\Pr_{x \in \{0,1\}^n} [ f(F^-(f, \text{unary}(n), f(x))) = f(x) ] < n^{-c}$$

where:

1. F⁻ is a PPT algorithm that attempts to find the inverse of f(x)
2. $n = \lceil \log_2(x) \rceil$ (the number of bits in x)
3. unary(n) is an all-ones bit string of size n, used by F⁻ to set an upper bound in its search
4. c is any positive integer

Intuitively, the asymptotic value of Pr as a function of n, will always be less than the asymptotic value of any polynomial of n, as n tends to infinity.

## 3. The Tau (τ) Functions Class

We define the class of functions T (Greek Tau) as the set of functions

$$T = \{ \tau \mid \tau : \{0,1\}^n \rightarrow \{0,1\}^n \},$$

where each τ is constructed as follows:

1. Construct a sequence M of n matrices $M=(M_0, M_1, \ldots, M_n)$, each one of size n x n, where each element is a random bit, bits in each matrix are uniformly distributed, and each $M_i \in M$ is unique.
2. Pick randomly a sequence of n hash functions $H\tau_{row} = (h_{row1}, h_{row2}..h_{rown})$, from a universal hash family: $\{ h : \{0,1\}^n \rightarrow \{0,1\}^{\log(n)} \}$
3. Pick randomly a sequence of n hash functions $H\tau_{col} = (h_{col1}, h_{col2}..h_{coln})$, from a universal hash family: $\{ h : \{0,1\}^n \rightarrow \{0,1\}^{\log(n)} \}$
4. Construct an n x n matrix $H_M$ of hash functions by picking randomly from a universal hash family: $\{ h : \{0,1\}^n \rightarrow \{0,1\}^3 \}$

5. Construct an 8 x 2 matrix D with the following values, each of which represents a moving direction along coordinates in a matrix in M:

| | | |
|---|---|---|
| 1 | 0 | (up) |
| -1 | 0 | (down) |
| 0 | -1 | (left) |
| 0 | 1 | (right) |
| -1 | -1 | (up,left) |
| -1 | 1 | (up,right) |
| 1 | -1 | (down,left) |
| 1 | 1 | (down,right) |

All hash functions constructed for τ follow Carter and Wegman's construction [CW77] of the form:

$$((ax + b) \mod p) \mod t$$

where:
1. p is an n-bit prime number chosen at random, with the condition $p > 2$
2. a is an integer chosen at random, with the condition $0 < a < p$
3. b is an integer chosen at random, with the condition $0 < b_i < p$
4. t=n for all the members of $H\tau_{row}$ and $H\tau_{col}$; and t=8 for all the members of $H_M$

Each $h_i$ in any collection (sequence or matrix) $C_H$ of universal hash functions in τ is subject to the following constraints:
1. Each $h_i$ should be unique within $C_H$. That is, for any two distinct $h_i, h_j \in C_H$, $a_i \neq a_j$, $b_i \neq b_j$ and $p_i \neq p_j$.
2. Each $h_i$ should be independent from any other $h_j$ in $C_H$. That is, given two distinct $h_i, h_j$ in $C_H$, $(a_i x + b_i) \neq k(a_j x + b_j)$ for any integer k>0.

Once constructed, each τ maps an input x in $\{0,1\}^n$ to an output y in $\{0,1\}^n$ by the following algorithm $A_\tau$.

1. Let the output y initially be an empty binary sequence
2. For each $i \in (1,2,\ldots,n)$:
    a. Let $hr_i$ be the $i^{th}$ hash function in $H\tau_{row}$
    b. Let $hc_i$ be the $i^{th}$ hash function in $H\tau_{col}$
    c. Let $r = hr_i(x)+1$
    d. Let $c = hc_i(x)+1$
    e. Let $M_i$ be the ith matrix in the bit-matrix sequence M
    f. Let (r,c) represent a (row,column) coordinate in $M_i$
    g. For each $j = (1,2,\ldots n)$
        i. Let $hd_{(i,j)}$ be the hash function from the matrix $H_M$
        ii. Let $d = hd_{(i,j)}(x)+1$
        iii. Let $r_\Delta = D_{(d,1)}$
        iv. Let $c_\Delta = D_{(d,2)}$
        v. Let $r = r + r_\Delta$
        vi. Let $c = c + c_\Delta$
        vii. If either r,c are less than one, let it be equal to n
        viii. If either r,c are greater than n, let it be equal to 1
        ix. Let $b = M_{i(r,c)}$ (the bit at the coordinate (r,c) in $M_i$)
        x. Append b to the output y
3. $\tau(x) = y$

**Lemma 1:** For every $\tau$ in T, the algorithm $A_\tau$ computes $\tau$ in polynomial time.
**Proof:**
1. All the variable assignment and hashing steps are O(1)
2. Step 2 computes y in $O(n^2)$, since:
    a. The outer loop in Step 2 is linear in n, thus it is O(n).
    b. Step 2.g is linear in n, thus it is O(n)
    c. The complexity of the outer loop in terms of the inner loop is O(n times (n + k)) for some k >1, so it is $O(n^2)$.
3. The sum of the complexities of the sequential steps is $O(n^2)$

## 4. Related Works

The construction of one-way functions has been the subject of extensive research in the past three decades, especially with applications in cryptography [L03]. Most of the possible-one-way function constructions for cryptographic applications are based on the assumption of the existence of one-way functions. Good surveys of application of one-way hash functions in cryptography can be found in [IZM90] and [NY89].

Our work is closely related to the use of random predicates [G00] and expander graphs [CLG09] in the use of a random walk along a graph to generate the value of the output of a hash function. One obvious difference is that our proposed functions are not hash functions. Disregarding this fact, our work is related to the use of random predicates in the sense that the output of the function is a composition of bits generated from random predicates; however, our constructions are based on random walks using the output of several unique hash functions, which avoids the need to use a lookup table to store the mappings of random predicates. In the case of expander graphs, the random walk is made along a set of vertices, each of which represents an elliptic curve, and the "current" input bit is used to decide between two isogenous curves. The output of the hash is a function of the j-invariant of the last vertex of the random walk. The one-wayness of functions based on this kind of expander graphs depends on the assumption of the hardness of finding isogeneous elliptic curves, which is an NP-complete problem; thus, the proof of one-wayness is dependent on the resolution of the P vs. NP problem.

Our work is also related to previous work based on composition of universal hash functions. In [SV00] such compositions are intended to build one-way cryptographic hash functions that break a message into independent sequences for which a set of randomly-picked universal hash functions are applied and whose output concatenated to produce the hashed output. In [BP97], the composition is made by breaking the input into fixed-sized strings to which distinct hash functions are applied, each of whose output is used to compose the function's output by application of other hashes to the outputs in cascade (either linearly or from a tree), being the final output the result of the hash function in the last stage. Notice that in neither of these works, traversals through a sequence of distinct random bit matrices are used.

## 5. The Tau-Inverse (T-INV) problem

We define T-INV as the problem of finding the inverse of a member $\tau$ of T, given a member y of the image of $\tau$. More formally, we define it as:

T-INV($\tau$,y) = { x | y = $\tau$(x) , $\tau \in$ T }

For the sake of consistency and to avoid repeating definitions in the rest of this paper, we will define the following. We shall use y to represent a member of the image of $\tau$, and define it as a sequence of bits $(y_1,\ldots,y_n)$, where n is the number of bits, and $y_1$ is the left-most bit (this last definition is arbitrary and is used only for visualization purposes).

We define H as the set of universal hash functions used to build $\tau$. More formally, $H = H\tau_{row} \cup H\tau_{col} \cup H_M$. We shall use $h_i$ to designate a member of H, where i is the index of the member, corresponding to the $i^{th}$ bit of y. When necessary, we shall clarify the context in which $h_i$ is used, or as a more specific member on either $H\tau_{row}$, $H\tau_{col}$ or $H_M$.

To simplify our analysis, when appropriate, we shall make O(1) oracle assumptions when establishing bounds in time complexities. Such assumptions are pertinent when proving exponential bounds, as constant and polynomial components are negligible as an exponential function grows asymptotically.

## 6. The H-Inverse (H-INV) problem

We define H-INV as the problem of finding *all* the members of the preimage of $h_i^-$(m), where $h_i$ is a member of H, as previously defined, and m is a member of the image of $h_i$. More formally, we define it as:

H-INV($h_i$,m) = { x | m = $h_i$(x) , $h_i \in$ H }

**Lemma 2:** Let M be the image of $h_i \in$ H and $|M| = 2^t$ for some t > 0. Any algorithm $A^-_H$(hi,m) that solves H-INV for $h_i$ and m $\in$ M has a lower-bound time complexity bounded by $\Omega(2^n)$.

**Proof:** Due to the universal nature of $h_i$, there is a guarantee that all members of its preimage are uniformly mapped to the members of its image. Thus, for any of its outputs, the size of the preimage is $2^n/2^t = 2^{n-t}$. The size of the output of $A^-_H$ will be $2^{n-t}$, thus the execution time to produce the output is bounded by $\Omega(2^n)$.

**Corollary 1:** Any algorithm that solves T-INV by invoking H-INV will have a best-case execution exponential time, bounded by $\Omega(2^n)$.

## 7. Reducing T-INV to FSAT is not in P

Reducing the problem of finding the inverse of a one-way function candidate to formula satisfiability (FSAT) would be a means of solving T-INV in polynomial time, if one can prove that P=NP. Since any $\tau$ in T can be implemented by a polynomial-size circuit, there exist a Boolean formula $\varphi$ that represents $\tau$. However, a polynomial-size circuit does not guarantee a polynomial size for $\varphi$; in fact, the immense majority of Boolean functions have a formula size that is exponential in n [S38].

We shall construct our proofs using the following definitions:
1. $R\tau(x,y)$ is a relation that represents a $\tau$ in T, such that $<x,y> \in R\tau$ if and only if $\tau(x) = y$.
2. The expression xy is the concatenation of a pair $<x,y>$ in $R\tau$, where both x and y are zero-padded to the left to ensure that their length is n.
3. The expression xy $\oplus$ x'y' represents a bit-wise XOR operation of xy and x'y'.
4. The function OC(*s*) is defined as OC : $\{0,1\}^* \to N$ and computes the number of ones in the bit string s.
5. A predicate IR(xy) (meaning that xy is *irreducible*) is true if there does not exist $<x',y'>$ in $R\tau$ such that x'y' $\neq$ xy and OC(xy $\oplus$ x'y') = 1.
6. The set $S_{IR}(\tau)$ is defined as $S_{IR}(\tau) = \{ xy \mid <x,y> \in R\tau$ and $IR(xy) \}$

**Lemma 3:** There exist instances $\tau$ in T for which $|S_{IR}(\tau)| > 2^n/k$, where k is an odd integer constant and $2 < k < n$.

**Proof:** Each $\tau$ is a *total function*, i.e. is defined for all members of its domain. Thus, there are $2^n$ members in $R\tau(x,y)$. The domain of $\tau$ contains values x = tk, where $t=\{0\ldots(2^n-1)/k\}$. Each one of such values can map to a value $y=u(k+2)$, where $u=\{1\ldots(2^n-1)/(k+2)\}$ (such mapping does not have to be one-to-one and in any specific order). For the values of $x \neq tk$, as defined above, each one can map to a value $y=w(k+4)$, where $w=\{1..(2^n-1)/(k+4)\}$ (once again, such mapping does not have to be one-to-one and in any specific order). Such a distribution of values guarantees that the predicate IR(xy) is true for at least $2^n/k$ mappings, thus $|S_{IR}(\tau)| > 2^n/k$.

**Lemma 4:** Reducing T-INV to FSAT is not in P.

**Proof:** We proved in Lemma 3 that there exist instances in $\tau$ in T for which $|S_{IR}(\tau)| > 2^n/k$. Mapping such instances to a Boolean formula in CNF results in an exponential number of irreducible clauses, as each of such clauses correspond to a member of $S_{IR}(\tau)$. Any TM that produces the formula will spend an exponential time producing the output. For a problem to be in P, *all instances* must have an algorithm that solves it in polynomial time. Thus, reducing T-INV to FSAT is not in P.

## 8. Worst-Case Lower Bound for Computing T-INV

We can always convert a deterministic Turing Machine (TM) into a Probabilistic TM where two random transitions from the same state bring the same output and have the same next state in common. Thus, we want to prove that solving the inverse of $\tau$ is not in FP, thus justifying the need of a PPT algorithm to compute the inverse in average polynomial time.

**Lemma 5:** Any deterministic algorithm $A\tau^-$ that attempts to find a member of the inverse of a function $\tau \in T$, has a worst-case execution time bounded by $\Omega(2^n)$.

**Proof:** Let $h_i$ be any of the hash functions used to construct $\tau$. For the purpose of analysis, let $A_O(\tau,w_i,j)$ be an O(1) oracle that finds the $j^{th}$ element of a sequence of members of the preimage $h_i^-(w_i)$. Let $y = \tau(x)$ for some x in the domain of $\tau$. Let $y_i$ be the $i^{th}$ bit of y. Each $y_i$ is produced by a bit from the corresponding matrix $M_i$, which resulted from the last step of a traversal, given by the value $w_i \in \{0..7\}$ of a hash function $h_i$ in $\tau$. Since $A\tau^-$ is deterministic, it has to pick one of eight possible hash values that lead to $y_i$ and invoke the oracle $A_O(\tau,w_i,j)$, where $wi \in \{0..7\}$ and $j \in \{1..2^n/8\}$ (as there can be as many as $2^n/8$ elements in the preimage $h_i^-$ due to the uniform distribution resulting from the universality of $h_i$). To find a member x of the inverse $\tau^-(y)$, $A\tau^-$ will need to find a value of x that produces a $y_i$ in y. In the worst case, such an x will be obtained by the $j=2^n/8$ call to the oracle $A_O(\tau,w_i,j)$. Thus, the worst-case execution time of $A\tau^-$ has a lower bound of $\Omega(2^n)$.

**Corollary 2:** T-INV is not in FP.

## 9. All members of τ are One-Way Functions

In this section, we shall prove that all members of T are one-way functions.

**Lemma 6:** Let $y = \tau(x)$ for some $x \in \{0,1\}^n$. Let $H_M$ be the matrix of universal functions used to construct $\tau$, and let $H_{Mi}$ be the $i^{th}$ row of the $H_M$. Let $h_i \in H_{Mi}$, where $i \in \{1,2,\ldots n\}$, let $y_i$ be the $i^{th}$ bit of y. Let $(d_{i1},d_{i2},\ldots,d_{in})$ be the output of each respective $h_i$ for an x in the domain of $\tau$, which defined the path to $y_i$. Let $F_i$ be the event of finding at random one element from the preimage $\tau^-(\{y\})$ with a path to $y_i$. The probability of $F_i$ is given by:

$$Pr[F_i] = 1/8^n$$

**Proof:** Since $h_i$ in $H_M$ is member of a universal family of hash functions, it is guaranteed that the size of the preimage $|h_i^-(\{y_i\})| = 2^n/8$. The size of the domain of $h_i$ is $2^n$, thus the probability of finding an x in the preimage $h_i^-(\{y_i\})$ is $2^n/(2^n/8) = 1/8$. Since all hash functions are independent, the probability of finding an x common to all $h_i$ is given by:

$$Pr[F_i] = \prod_i (Pr[x \in h_i^-(\{y_i\})]) = \prod_i (1/8) = 1/8^n \qquad i = \{1,2,\ldots n\}$$

**Lemma 7:** Let $y = \tau(x)$, $y_i$, $H_M$, $H_{Mi}$ and $h_i$ be as defined in Lemma 6. Let $F_i$ be the event of finding at random one element from the preimage $\tau^-(\{y\})$ with a path to $y_i$. Let $F_j$ be the event of finding at random one element from the preimage $\tau^-(\{y\})$ with a path to $y_j$, and $i \neq j$. The probability of $F_i$ given $F_j$, when $F_i \neq F_j$, is given by:

$$Pr[F_i \mid F_j] = \alpha \cdot 1/8^n$$

for some constant $0 < \alpha < 1$.

**Proof:** We know that:

$$Pr[F_i \mid F_j] = Pr[F_i \cap F_j] / Pr[F_j]$$

and $\Pr[F_k] = 1/8^n$ for all $k = \{1,\ldots,n\}$, and that $F_j \not\subseteq F_i$. Since we are assuming some degree of dependency between $F_i$ and $F_j$ (otherwise the conditional probability would be zero), it follows that

$$\Pr[F_i \cap F_j] = \alpha \cdot 1/8^n \cdot 1/8^n$$

for some constant $0 < \alpha < 1$. Replacing in the equation yields to:

$$\Pr[F_i \mid F_j] = \alpha \cdot 1/8^n$$

**Lemma 8:** Let $y = \tau(x)$ for some $x$ in the domain of $\tau$. Let $F_i$ be the event of finding at random one element from the preimage $h_i^-(\{y_i\})$. Let $F_{-i}$ be intersection:

$$F_{-i} = \bigcap_{k \neq i} F_k$$

The probability of $F_{-i}$ is given by:

$$\Pr[F_{-i}] = \beta \cdot (1/8^{(n^2-n)})$$

for some constant $0 < \beta < 1$.

**Proof:** From Lemma 6, we know that $\Pr[F_k] = 1/8^n$ for all $k = \{1,\ldots,n\}$. Since we are assuming some dependency among events, the probability of the intersection is given by:

$$\Pr[F_{-i}] = \beta \prod_{k \neq i} (\Pr[F_k])$$

for some constant $0 < \beta < 1$. Thus,

$$\Pr[F_{-i}] = \beta \cdot (1/8^n)^{(n-1)}$$

which is equivalent to

$$\Pr[F_{-i}] = \beta \cdot (1/8^{(n^2-n)})$$

**Lemma 9:** Let $F_i$ and $F_{-i}$ be as defined in Lemmas 6 and 7. The probability of $F_i$ given $F_{-i}$ is given by:

$$\Pr[F_i \mid F_{-i}] = \gamma \cdot 1/8^n$$

**Proof:** We know that

$$\Pr[F_i \mid F_{-i}] = \Pr[F_i \cap F_{-i}] / \Pr[F_{-i}]$$

from Lemma 7.

$$\Pr[F_i \cap F_{-i}] = \alpha \cdot 1/8^n \cdot 1/8^{(n^2-n)}$$

for some constant $0 < \alpha < 1$. From Lemma 8:

$$\Pr[F_{-i}] = \beta \cdot (1/8^{(n^2-n)})$$

Replacing in the formula above gives:

$$\Pr[F_i \mid F_{-i}] = \alpha \cdot 1/8^n \cdot 1/8^{(n^2-n)} / (\beta \cdot (1/8^{(n^2-n)}))$$

which brings:

$$\Pr[F_i \mid F_{-i}] = \gamma \cdot 1/8^n$$

where $\gamma = \alpha / \beta$ and $0 < \gamma < 1$.

**Lemma 10:** Let y, $y_i$ and $F_i$ be as defined in Lemma 6. Let F be the event of finding at random one member of the preimage $\tau^-(\{y\})$. The probability of F is given by:

$$\Pr[F] = \Pr[\bigcap_i F_i]$$

**Proof:** Each $y_i$ is produced independently by a set of x in the preimage $\tau^-(\{y\})$. However, due to the universal-hash guarantee and the distinct parameters of each hash function in $\tau$, there exists an x that produces a bit in y but does not produce same bit value in another bit of y. Thus, the preimage $\tau^-(\{y\})$ is given by the intersection of the independent preimages for each bit $y_i$ in y.

With these lemmas proved, we are now in a position to prove that all members of T are one-way functions.

**Theorem 1:** Let $y = \tau(x)$ for some $x \in \{0,1\}^n$. Let F be the event of finding at random one member of the preimage $\tau^-(\{y\})$. The probability of F is bounded by:

$$\Pr[F] < n^{-c}$$

where c is any positive integer.

**Proof:** From Lemma 10, we know that the probability of F is given by:

$$\Pr[F] = \Pr[\bigcap_i F_i] \qquad\qquad i \in \{1,2,\ldots n\}$$

which is equivalent to:

$$\Pr[F] = \Pr[F_1 \cap F_2 \ldots \cap F_n]$$

Given two distinct $F_i$ and $F_j$, the probability of their intersection is given by:

$$\Pr[F_i \cap F_j] = \Pr[F_i \mid F_j] \Pr[F_j]$$

By the chain rule for multiple intersections of sets, the probability of F in terms of the intersection of all $F_i$ is given by:

$$\Pr[F] = \Pr[\bigcap_i F_i] = \Pr[F_n \mid F_{n-1} \cap F_{n-2} \ldots F_1] \cdot \Pr[F_{n-1} \mid F_{n-2} \cap F_{n-3} \ldots F_1] \ldots \cdot \Pr[F_1]$$

Since $\Pr[F_i] = 1/8^n$ for any $F_i$ (Lemma 6), then:

$$\Pr[\cap_i F_i] = \Pr[F_n \mid F_{n-1} \cap F_{n-2} \ldots F_1] \cdot \Pr[F_{n-1} \mid F_{n-2} \cap F_{n-3} \ldots F_1] \ldots \cdot 1/8^n$$

We proved in Lemma 7 that $\Pr[F_i \mid F_{-i}] = \gamma \cdot 1/8^n$ for some constant $0 < \gamma \leq 1$ and any $F_{-i}$ that is the intersection of all $F_j \neq F_i$. Thus, $\Pr[F_i \mid F_{-i}] = 1/\beta_i$ for all $F_i$, for some constant $\beta_i > 1$. Replacing each term in the equation above with its equivalent $1/\beta_i$:

$$\Pr[\cap_i F_i] = 1/8^n \prod_i (1/\beta_i) \qquad i = \{1,2,\ldots,n\}$$

Since each term in the $\Pi$ operation is less than one, then:

$$\Pr[\cap_i F_i] \leq 1/\alpha^n \qquad i = \{1,2,\ldots,n\}$$

where $\alpha = \max(\beta_1, \beta_2, \ldots, \beta_n)$. Therefore,

$$\Pr[F] \leq 1/\alpha^n$$

for some constant $\alpha > 1$. Since $\Pr[F]$ is inversely exponential in n, it follows that:

$$\Pr[F] < n^{-c}$$

**Corollary 3:** Each $\tau \in T$ is a one-way function.

## 10. Conclusion

We have proved that $P \neq NP$ presented with the proof of the existence of a class of functions T, each of whose members satisfies the conditions of one-way functions. Each member in T is constructed with a collection of independent universal hash functions that produce a starting coordinate and path within a sequence of unique bit matrices. A proof of the exponential lower bound of worst-case complexity of finding the inverse of members of T was presented. We also proved that the problem of finding the preimage of each member in T is not polynomial-time reducible to FSAT. It was also proved that any random algorithm that attempts to find the inverse of any function in T has negligible probability of success, thus proving that all members of T are one-way functions.